\documentclass[screen,sigconf]{acmart}
\AtBeginDocument{%
  }

\copyrightyear{2026}
\acmYear{2026}
\setcopyright{cc}
\setcctype{by}
\acmConference[LAK 2026]{LAK26: 16th International Learning Analytics and Knowledge Conference}{April 27-May 01, 2026}{Bergen, Norway}
\acmBooktitle{LAK26: 16th International Learning Analytics and Knowledge Conference (LAK 2026), April 27-May 01, 2026, Bergen, Norway}
\acmDOI{10.1145/3785022.3785026}
\acmISBN{979-8-4007-2066-6/2026/04}

\usepackage[capitalise,noabbrev]{cleveref}
\usepackage{bm}
\usepackage{algorithm}
\usepackage{algorithmic}
\usepackage{subcaption}
\usepackage{multirow}
\usepackage{booktabs}
\usepackage{xcolor}

\begin{document}

\def\gA{{\mathcal{A}}}
\def\gB{{\mathcal{B}}}
\def\gC{{\mathcal{C}}}
\def\gD{{\mathcal{D}}}
\def\gE{{\mathcal{E}}}
\def\gF{{\mathcal{F}}}
\def\gG{{\mathcal{G}}}
\def\gH{{\mathcal{H}}}
\def\gI{{\mathcal{I}}}
\def\gJ{{\mathcal{J}}}
\def\gK{{\mathcal{K}}}
\def\gL{{\mathcal{L}}}
\def\gM{{\mathcal{M}}}
\def\gN{{\mathcal{N}}}
\def\gO{{\mathcal{O}}}
\def\gP{{\mathcal{P}}}
\def\gQ{{\mathcal{Q}}}
\def\gR{{\mathcal{R}}}
\def\gS{{\mathcal{S}}}
\def\gT{{\mathcal{T}}}
\def\gU{{\mathcal{U}}}
\def\gV{{\mathcal{V}}}
\def\gW{{\mathcal{W}}}
\def\gX{{\mathcal{X}}}
\def\gY{{\mathcal{Y}}}
\def\gZ{{\mathcal{Z}}}

\def\sA{{\mathbb{A}}}
\def\sB{{\mathbb{B}}}
\def\sC{{\mathbb{C}}}
\def\sD{{\mathbb{D}}}
\def\sE{{\mathbb{E}}}
\def\sF{{\mathbb{F}}}
\def\sG{{\mathbb{G}}}
\def\sH{{\mathbb{H}}}
\def\sI{{\mathbb{I}}}
\def\sJ{{\mathbb{J}}}
\def\sK{{\mathbb{K}}}
\def\sL{{\mathbb{L}}}
\def\sM{{\mathbb{M}}}
\def\sN{{\mathbb{N}}}
\def\sO{{\mathbb{O}}}
\def\sP{{\mathbb{P}}}
\def\sQ{{\mathbb{Q}}}
\def\sR{{\mathbb{R}}}
\def\sS{{\mathbb{S}}}
\def\sT{{\mathbb{T}}}
\def\sU{{\mathbb{U}}}
\def\sV{{\mathbb{V}}}
\def\sW{{\mathbb{W}}}
\def\sX{{\mathbb{X}}}
\def\sY{{\mathbb{Y}}}
\def\sZ{{\mathbb{Z}}}

\newcommand{\Range}{\mathrm{Range}}
\newcommand{\Proc}{\mathrm{Proc}}
\newcommand{\x}{\bm{x}}
\newcommand{\z}{\bm{z}}
\newcommand{\g}{\bm{g}}
\newcommand{\m}{\bm{m}}
\newcommand{\vb}{\bm{v}}
\newcommand{\thetab}{\bm{\theta}}
\newcommand{\phib}{\bm{\phi}}
\newcommand{\Cat}{\mathrm{Cat}}
\newcommand{\pib}{\bm{\pi}}
\newcommand{\pub}{\mathrm{pub}}
\newcommand{\priv}{\mathrm{priv}}
\newcommand{\syn}{\mathrm{syn}}
\newcommand{\mub}{\bm{\mu}}
\newcommand{\sigmab}{\bm{\sigma}}
\newcommand{\diag}{\mathrm{diag}}
\newcommand{\KL}{\mathrm{KL}}
\newcommand{\Mone}{\mathrm{M1}}
\newcommand{\AJS}{\mathrm{AJS}}
\newcommand{\JS}{\mathrm{JS}}
\newcommand{\recon}{\mathrm{recon}}
\newcommand{\gen}{\mathrm{gen}}

\title[Cyclic Adaptive Private Synthesis for Sharing Real-World Data in Education]{Cyclic Adaptive Private Synthesis for Sharing Real-World Data in Education}

\author{Hibiki Ito}
\email{hibiki.itoo@gmail.com}
\orcid{0000-0002-1558-8818}
\affiliation{%
  \institution{Kyoto University}
  \department{School of Informatics}
  \city{Kyoto}
  \country{Japan}
}

\author{Chia-Yu Hsu}
\email{hsu.chiayu.2u@kyoto-u.ac.jp}
\orcid{0000-0003-3581-4090}
\affiliation{%
  \institution{Kyoto University}
  \department{Academic Center for Computing and Media Studies}
  \city{Kyoto}
  \country{Japan}
}

\author{Hiroaki Ogata}
\email{hiroaki.ogata@gmail.com}
\orcid{0000-0001-5216-1576}
\affiliation{%
  \institution{Kyoto University}
  \department{Academic Center for Computing and Media Studies}
  \city{Kyoto}
  \country{Japan}
}

\renewcommand{\shortauthors}{Ito et al.}

\begin{abstract}
  The rapid adoption of digital technologies has greatly increased the volume of real-world data (RWD) in education.
  While these data offer significant opportunities for advancing learning analytics (LA), secondary use for research is constrained by privacy concerns.
  Differentially private synthetic data generation is regarded as the gold-standard approach to sharing sensitive data, yet studies on the private synthesis of educational data remain very scarce and rely predominantly on large, low-dimensional open datasets.
  Educational RWD, however, are typically high-dimensional and small in sample size, leaving the potential of private synthesis underexplored.
  Moreover, because educational practice is inherently iterative, data sharing is continual rather than one-off, making a traditional one-shot synthesis approach suboptimal.
  To address these challenges, we propose the Cyclic Adaptive Private Synthesis (CAPS) framework and evaluate it on authentic RWD.
  By iteratively sharing RWD, CAPS not only fosters open science, but also offers rich opportunities of design-based research (DBR), thereby amplifying the impact of LA.
  Our case study using actual RWD demonstrates that CAPS outperforms a one-shot baseline while highlighting challenges that warrant further investigation.
  Overall, this work offers a crucial first step towards privacy-preserving sharing of educational RWD and expands the possibilities for open science and DBR in LA.
\end{abstract}

\begin{CCSXML}
<ccs2012>
   <concept>
       <concept_id>10010405.10010489</concept_id>
       <concept_desc>Applied computing~Education</concept_desc>
       <concept_significance>500</concept_significance>
       </concept>
   <concept>
       <concept_id>10002978.10003029.10011150</concept_id>
       <concept_desc>Security and privacy~Privacy protections</concept_desc>
       <concept_significance>500</concept_significance>
       </concept>
   <concept>
       <concept_id>10010147.10010257</concept_id>
       <concept_desc>Computing methodologies~Machine learning</concept_desc>
       <concept_significance>300</concept_significance>
       </concept>
 </ccs2012>
\end{CCSXML}

\ccsdesc[500]{Applied computing~Education}
\ccsdesc[500]{Security and privacy~Privacy protections}
\ccsdesc[300]{Computing methodologies~Machine learning}

\keywords{Real-World Data, Data Sharing, Differential Privacy, Synthetic Data, Learning Analytics}


\maketitle

\section{Introduction}\label{sec:intro}

With the widespread integration of digital technologies, the last few decades have witnessed a considerable growth in the volume of real-world data (RWD) \cite{Mahajan2015RWD} in the realm of education.
Unlike data collected primarily for experimental research, RWD offers rich opportunities for advancing learning analytics (LA) by complementing experimental data and enabling the discovery of real-world evidence (RWE) \cite{Kuromiya2023RWD,Okumura2026RWE}.
However, access to sensitive educational data---such as digital trace data---remains restricted to trusted researchers and are seldom shared with the broader research community \cite{Baker2024open}.
These data enclaves slow the progress of LA and undermine open science, limiting the field's impact \cite{Baker2024open,Haim2023open}.

Sharing data for the public interest while protecting privacy is much easier said than done. 
Although no perfect solution exists, private synthesis---synthetic data generation under differential privacy (DP, \cite{Dwork2006DP}) guarantee---is considered to be the gold-standard approach due to its theoretical advantages in privacy protection \cite{Gadotti2024survey}.
A model, often deep learning–based, is trained to capture the statistical properties of the original data and then used to release either the model itself or the generated synthetic data.
DP provides a provable guarantee of information-theoretic privacy during the training procedure and is often regarded as offering sufficient privacy for sharing sensitive data when DP parameters are suitably calibrated \cite{Pilgram2025consensus}.

Despite its promise, applications of private synthesis to educational data have been very scarce.
Existing studies typically use publicly available open datasets that are large and low-dimensional \cite{Liu2025DP-SDG,Liu2025fairDP,Kesgin2025FairSYN}, while most educational contexts produce small, locally gathered datasets distributed across disparate platforms \cite{Nguyen2024smallLA}.
Multimodal and longitudinal data collection further raises dimensionality \cite{Mohammadi2025MMLA}.
This combination of small sample sizes and high dimensionality---a well-known challenge in the DP literature---makes private synthesis especially difficult for educational RWD \cite{Heine2023small}.
Moreover, most research assumes a one-shot setting, training a new generative model from scratch for each dataset \cite{Bun2024continual}.
Educational practice, however, is inherently cyclical: similar RWD arrive repeatedly across cohorts, making a one-off synthesis approach suboptimal.

To address these challenges, we introduce the Cyclic Adaptive Private Synthesis (CAPS) framework and evaluate it with authentic RWD in education.
CAPS exploits the regular arrival of comparable datasets---for example, yearly cohorts---by pre-training a feature extractor to handle small-sample, high-dimensional data and adapting it iteratively over time, thereby tailoring traditional one-shot synthesis techniques to real-world educational settings.
CAPS allows LA researchers to access RWD and possibly share it with the broader research community, thereby promoting open science.
In addition, a critical implication of CAPS is that the iterative process enables design-based research (DBR) in real-world education \cite{Collins1992DBR,Brown1992DBR}.
That is, by iteratively sharing RWD, LA researchers would be able to go through the cycle of analysing the data to provide usable insights to practitioners as well as developing theoretical knowledge \cite{Reimann2016LA-DBR}.
Hence, CAPS can significantly advance the field of LA through fostering open science and offering rich opportunities of DBR for LA researchers.

Our case study using RWD from a secondary-school mathematics class demonstrates that the model utility evaluated by downstream classification performance and reconstruction power improve over successive cycles.
This indicates that CAPS allows the generative model to effectively learn statistical properties of sensitive RWD, outperforming the traditional one-shot baseline.
Yet, careful analysis also reveals that the quality of synthetic data could slightly degrade according to our evaluation metric.
We term this phenomenon the \emph{compounding bias effect}, indicating a potential area of concern that warrants further investigation.
Overall, this paper takes a crucial first step towards sharing RWD in education and thereby significantly increasing the impact of LA.

\subsection{Related works}\label{sec:related-works}

While various privacy protection techniques have been studied for sharing educational RWD \cite{Liu2023review}, this paper particularly focuses on the application of DP and synthetic data in LA. Gursoy et al. \cite{Gursoy2017DPLA} first demonstrated the potential of DP in LA, inspiring subsequent applications such as grade prediction \cite{Zhao2025DPpredict} and knowledge tracing \cite{Kabir2025DPKT}.
Broader frameworks for incorporating DP into LA have also been proposed \cite{Liu2025DPframework,Podsevalov2022DPframework}.
These, however, focus on privacy-preserving predictive tasks and data analysis rather than data sharing.
Private synthesis becomes essential when sensitive RWD must be shared within the research community while allowing for various downstream tasks, yet it has received little attention in education.
Notable studies include those of Liu et al.\ \cite{Liu2025DP-SDG,Liu2025fairDP}, which tested private aggregation of teacher ensembles (PATE) frameworks and generative adversarial networks (GAN)-based methods, and the work by Kesgin \cite{Kesgin2025FairSYN}, which examined a private diffusion-based model.
However, these deep learning models typically require very large datasets, and training them with DP on small high-dimensional data has proved practically infeasible without public auxiliary information such as pre-training \cite{Ganesh2023pretrain,Ben-David2023compression}.
Moreover, open datasets used in these prior works are large and low-dimensional, differing substantially from the sensitive, small-scale RWD that ultimately need to be shared.

Existing private-synthesis research also assumes a one-shot paradigm: a model is trained anew for each dataset release.
Our work instead targets \emph{iterative} data sharing.
The proposed cyclic synthesis should not be confused with the emerging \emph{longitudinal} synthesis in the DP literature, which divides longitudinal datasets---such as census data---into temporal segments for continual release \cite{Bun2024continual,He2024online}.
Those approaches aim to repeatedly release data from the same individuals, whereas CAPS generates synthetic data across successive, distinct cohorts while retaining consistent educational contexts.
This latter perspective also enables cyclic interventions and the development of usable theoretical knowledge through a DBR approach.

In summary, the contribution of this paper is twofold. 
First, we present the CAPS framework, which aims to optimise private synthesis for iterative sharing of educational RWD while addressing both small-sample and high-dimensional challenges.
Second, we validate CAPS on authentic RWD that are small and longitudinal (i.e.\ high-dimensional) demonstrating its effectiveness in realistic educational settings.

% For example, Bun et al.\ \cite{Bun2024continual} formalised this problem by a theoretical database of binary bits and illustrated how census-like longitudinal data can be continually released.

\section{Cyclic Adaptive Private Synthesis (CAPS) framework}\label{sec:framework}

We first delineate a few preliminary definitions regarding DP and the model of Kingma et al.\ \cite{Kingma2014SSL} which is a core of our framework.
Subsequently, the CAPS framework is described based on these definitions.

\subsection{Preliminary (1): differential privacy}\label{sec:dp}

We employ the standard approximate DP defined as follows: we say that datasets $D$ and $D^\prime$ are adjacent datasets if they differ in a single data point by addition or removal.

\begin{definition}[Differential privacy \cite{Dwork2006approxDP}]\label{def:dp}
    An algorithm $\gA$ is $(\varepsilon, \delta)$-differentially private if for all $\gS \subseteq \Range(\gA)$ and for all adjacent datasets $D$ and $D^\prime$:
    \begin{equation}
        \Pr(\gA(D) \in \gS) \leq e^\varepsilon \Pr(\gA(D^\prime) \in \gS) + \delta,
    \end{equation}
    where probabilities are over the randomness in the algorithm $\gA$.
\end{definition}

Here, a data point in $D$ is called a \emph{privacy unit} as it defines the adjacency. 
In the following, we assume that a privacy unit is a distinct individual learner (i.e.\ user-level DP).
However, our framework is flexible enough to allow for a relaxed privacy unit such as data for a certain time window.

The following important property of DP will be also used in our framework:

\begin{proposition}[Post-processing \cite{Dwork2014foundations}]\label{prop:postprocessing}
    If an algorithm $\gA$ satisfies $(\varepsilon, \delta)$-DP, then a post-processing $\Proc\circ\gA$ is also $(\varepsilon, \delta)$-DP.
\end{proposition}

Additionally, it is convenient to clarify the distinction between public and private data. Informally, incorporating public data to the computation of a private algorithm does not consume privacy budget. The following definition is adapted from Hod et al.\ \cite{Hod2025surrogate} and Ben-David et al.\ \cite{Ben-David2023compression}.

\begin{definition}[Public data]\label{def:piblic-data}
    A dataset $D^\prime$ is public if for an algorithm $\gA$ satisfying $(\varepsilon, \delta)$-DP and a private dataset $D$, both $\gA(D, \cdot)$ and $\gA(D, D^\prime)$ satisfy identical $(\varepsilon, \delta)$-DP guarantee.
\end{definition}

\subsection[Preliminary (2): model of Kingma et al.]{Preliminary (2): model of Kingma et al.\ \cite{Kingma2014SSL}}

We wish to train a generative model with DP for a small and high-dimensional dataset $D=\{(\x_i, y_i)\}_{i=1}^N$, where $\x_i \in \sX \subseteq \sR^d$ with dimension $d$ are features and $y_i \in \sY = \{1, 2,\dots, L\}$ are labels.
By far the most practical approach is to first pre-train a large model on public data and then fine-tune a small (additional) part of it on $D$ \cite{De2022DPscale,Tramer2022feature}.
Particularly, unlike the prior works, we employ the variational autoencoder (VAE) \cite{Kingma2014VAE,Rezende2014VAE} as a base model since it is relatively stable for small and high-dimensional data \cite{Mahmud2020VAEsmall}.
The following model introduced by Kingma et al.\ \cite{Kingma2014SSL} combines a pre-trained large VAE (called M1) and a small conditional generative model (called M2) trained by semi-supervised learning.
The separation of M1 and M2 allows for reusing the feature extractor across datasets of different label spaces.
Thus, we adopt it as the core model of our proposed framework.
In particular, we do not just reuse M1, but cyclically improve it over time (hence the name \emph{cyclic adaptive}).

Let $p(\x)$, $p(y)$ and $p(\z)$ denote the prior distributions over the feature variables $\x$, the label variable $y$ and the latent variables $\z$, respectively.
Following Kingma et al.\ \cite{Kingma2014SSL}, we formulate a probablistic model that consists of two models: the first model, M1, is an unconditional VAE with latent variables $\z_1$:
\begin{align}
    p(\z_1) &= \gN(\z_1; 0, I) \\
    p_{\thetab_1}(\x\mid \z_1) &= f_1(\x; \z_1, \thetab_1),
\end{align}
where $\gN(\cdot; 0, I)$ denotes the density of the standard normal distribution and $f_1(\x;\z_1, \thetab_1)$ is a suitable likelihood function with parameters $\thetab_1$.
To enable conditional generation, a smaller conditional variant of VAE, M2, is stacked on top of M1:
\begin{align}
    p(y) &= \Cat(y; \pib) \\
    p(\z_2) &= \gN(\z_2; 0, I); \\
    p_{\thetab_2}(\z_1\mid y, \z_2) &= f_2(\z_1; y, \z_2, \thetab_2),
\end{align}
where $\Cat$ denotes a categorical distribution parameterised by $\pib$ and $f_2(\z_1; y, \z_2, \thetab_2)$ is a suitable likelihood function.
Here, we assume that the priors of the latent variables $\z_1$ and $\z_2$ are Gaussians, but our framework is open to other variants such as vector quantised VAE \cite{van-den-Oord2017VQVAE}.

To train this M1+M2 stacked model, we first train M1 to learn latent variables $\z_1$ with large unlabelled data $\x$ by a standard VAE training \cite{Kingma2014VAE}. 
Subsequently, we freeze M1 and train M2 using latent representations derived from M1 in a semi-supervised manner, where the label variable $y$ is treated as a latent variable for unlabelled points.
As a result, we have the following probablistic model:
\begin{equation}
    p_{\thetab}(\x, y, \z_1, \z_2) = p(y)p(\z_2)p_{\thetab_2}(\z_1\mid y, \z_2)p_{\thetab_1}(\x\mid \z_1).
\end{equation}

\subsection{CAPS}\label{sec:caps}

Now we describe the Cyclic Adaptive Private Synthesis (CAPS) framework.
To grasp the idea, consider the following example setting.
Suppose that an LA system is deployed at an undergraduate study module.
Let $D_1 = \{(\x_i, y_i)\}_{i=1}^{N_1}$ be a dataset of $N_1$ students who participated in the module in year $t=1$ (or, more generally, cycle $t=1$), where $\x_i \in \sX$ are data obtained from the system and $y_i \in \sY_1$ are final exam scores. 
According to the feedback from students and data analytics, the instructor decides to replace the final exam by an essay assignment in the following year.
Let $D_2 = \{(\x_i, y_i)\}_{i=1}^{N_2}$ denote the dataset for year $t=2$.
Then the labels $y_i \in \sY_2$ should now contain the evaluation of the essay assignment, so the label spaces $\sY_1$ and $\sY_2$ are distinct.
We assume that the feature space $\sX$ remains the same (i.e.\ the data collection methods are the same) and that the distributions $p_1(\x)$ and $p_2(\x)$ over the features do not significantly differ (i.e.\ the cohorts are similar).
The above procedure is repeated for a few times, producing datasets $D_1, D_2, D_3, \dots$ with label spaces $\sY_1, \sY_2, \sY_3, \dots$.
Sharing such RWD allows researchers to discover RWE such as how different evaluation methods impact learning processes and generate hypotheses about, for example, how to improve the system to enhance learning.

\begin{figure*}[tb]
    \centering
    \includegraphics[trim=0cm 7.9cm 4.8cm 2.3cm, clip,width=\linewidth]{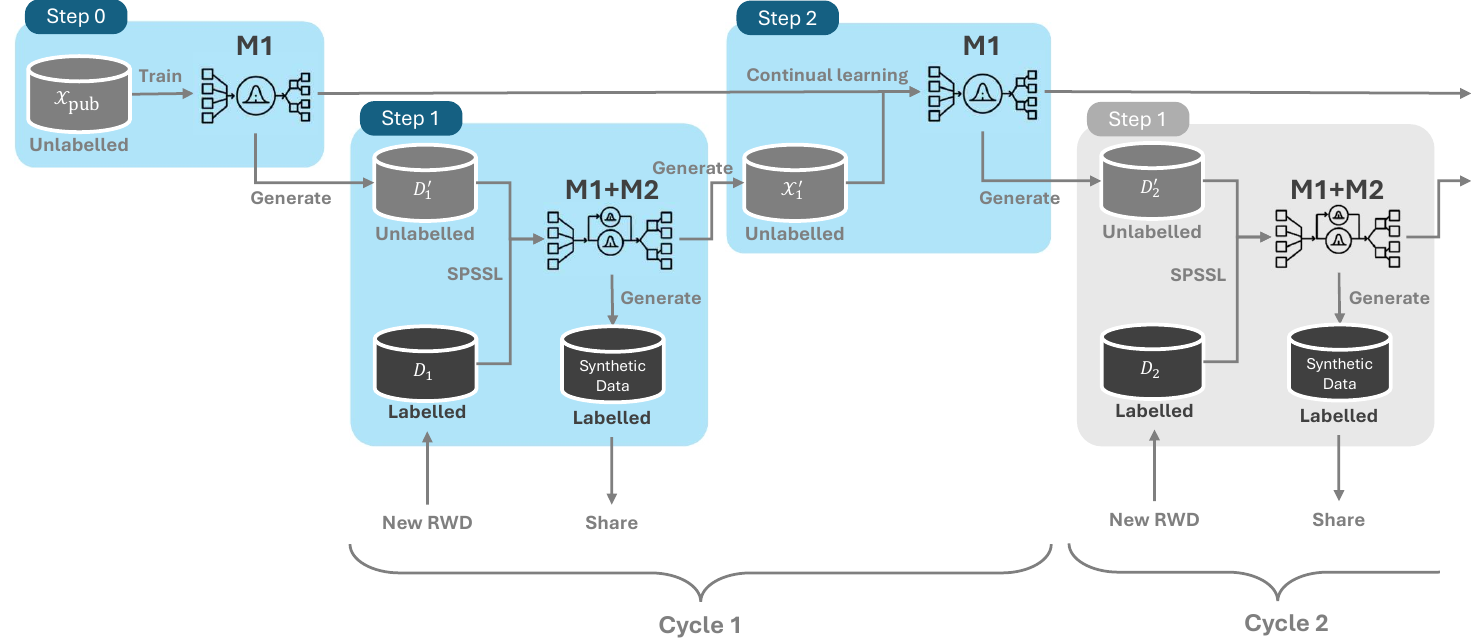}
    \Description{A diagram of the CAPS framework shows step 0, 1, and 2 for cycle 1 and continues to step 1 of cycle 2.}
    \caption{Overview of the proposed CAPS framework. $D_t$ are private datasets for cycles $t=1,2,\dots$ which we wish to share with third parties. The generative model M1+M2 is trained by semi-private semi-supervised learning (SPSSL) to share the synthetic data or the model itself under DP guarantee.}
    \label{fig:caps}
\end{figure*}

To generate synthetic data for these datasets using DP, our CAPS framework proceeds as follows (see \cref{fig:caps}).
Note that cycles do not have to be years (e.g.\ semesters) as long as we have distinct privacy units for the different cycles.

\begin{itemize}
    \item[\textbf{Step 0}] \emph{Pre-train M1}. We initialise CAPS by pre-training M1, a larger unconditional VAE, on large unlabelled public data $\gX_\pub$ whose feature space is the same as that of the private data.

    \item[\textbf{Step 1}] \emph{Train M2 for cycle $t$}. Given a pre-trained M1, we generate unlabelled data from it, denoted as $D_t^\prime$.
    Then M2, a smaller conditional generative model stacked on the current M1, is trained on $D_t \cup D_t^\prime$ by semi-private semi-supervised learning (SPSSL).
    To satisfy DP, SPSSL typically adds noise to a normal semi-supervised learning only when processing private data points \cite{Alon2019semi-private,Pinto2024PILLAR}.
    Note that $D_t^\prime$ satisfies \cref{def:piblic-data} in this training process, thereby regarded as public data.
    Consequently, since the output model M1+M2 satisfies DP, by \cref{prop:postprocessing}, we may share the trained model or synthetic data generated from it with third-party researchers.
    Now that we shared the data, we move on to Step 2 if there is cycle $t+1$.

    \item[\textbf{Step 2}] \emph{Update M1}.
    For some $n$, let $\gX_t^\prime=\{\x_i^\prime\}_{i=1}^n$ be synthetic features generated by the M1+M2 stacked mode just trained.
    It should be noted that $\gX_t^\prime$ can be treated as public for cycle $t+1$ by \cref{prop:postprocessing} and \cref{def:piblic-data}.
    We expect that teaching the private knowledge contained in $\gX_t^\prime$ to M1 will improve the prior for the subsequent cycles.
    Therefore, we update M1 using $\gX_t^\prime$.
    Note that simply fine-tuning M1 on $\gX_t^\prime$ would result in \emph{catastrophic forgetting} of previous training data that contain potentially useful information for the subsequent cycles \cite{French1999forget}.
    Hence we employ the approach of continual learning \cite{Wang2024CL}.
    Now that the M1 is updated, we go back to Step 1 with $t \leftarrow t + 1$.
\end{itemize}

\section{Case study: materials and methods}

In this section we instantiate the proposed CAPS framework with actual educational RWD as a case study.
We focus on learning habits study as an example LA research \cite{Shirvani-Boroujeni2019patterns,Hsu2024design,Ricker2020click}.
There has been evidence that forming a habit of learning---defined as a repetitive behaviour in the context of learning \cite{Wood2007habit}---has a significant effect on learning such as academic achievement \cite{Shirvani-Boroujeni2019patterns} and productivity \cite{Hsu2024productivity}.
Since learning habits data may allow for inferring daily routines of individual learners, it is very sensitive and individual privacy should be carefully protected when data are shared with third parties.
In this case study, we particularly focus on K-12 context.
As learning habits study typically involves longitudinal data collection and it is especially challenging to obtain large samples in K-12 context, this gives rise to the small-sample and high-dimensionality issues.

\subsection{Materials}

\subsubsection{Context}

RWD was obtained from a Japanese lower-secondary school over three years (2022-2024).
In the 7th-grade mathematics class of the school, students have a short practice test every week to check the understanding of learning contents.
The topic of each weekly test is announced beforehand and corresponding learning materials are given to the students on an e-book platform called BookRoll \cite{Ogata2015BR}.
The BookRoll system allows for collecting log data of students' interactions with the materials in the xAPI format, and the collected data are stored in learning record store (LRS).
The materials are not mandatory assignments, but students are encouraged to use them to prepare for the weekly tests.
To help students' self-directed learning, the goal-oriented active learning (GOAL) system \cite{Li2021GOAL} has been deployed at the school, on which students can manually enter weekly test scores by themselves and also monitor their own activity records such as time spent on studying. 

\subsubsection{Data}

For each year, we extracted from the LRS log data of the 7th-grade students' interactions with the practice materials over 17 weeks, corresponding to one semester of the school.
Additionally, end-of-semester exam scores of mathematics were obtained.
We only included to the datasets those who have at least one log record on the learning materials during these periods.
As a result, the datasets contain log data (features) and exam scores (labels) for 105, 111 and 115 students for years 2022, 2023 and 2024, respectively.
Although CAPS supports distinct label spaces, we retain identical ones across the three datasets for consistent evaluation.

\subsubsection{Pre-processing}

Now we wish to train generative models for these datasets.
However, using raw log data with granular timestamps and several features is infeasible, especially for such small samples due to the significant signal-to-noise ratio.
Though in practice it is often convenient to keep the data close to the raw form with minimum feature engineering so that third-party researchers could conduct a wider range of analyses, some pre-processing would be necessary for feasible private synthesis.
Indeed in this case study, we conduct extensive feature engineering to simplify the settings.

An important form of data in LA is time series, as it allows for exploring temporal changes within individuals and personalising learning \cite{Saqr2026within}.
Following the prior work by Hsu et al.\ \cite{Hsu2024productivity,Hsu2023chronotypes}, we first estimate time-on-task for each hour as the difference between the first and the last log record within the one-hour time window.
Then these are aggregated into four time frames of the day categorised by Ricker et al.\ \cite{Ricker2020click}: morning (05:00-11:59), afternoon (12:00-16:59), evening (17:00-23:59) and overnight (00:00-05:00).
To further simplify data, we aggregate time-on-task of each weak for each time frame into three engagement classes: inactive (zero minutes), active (1 to 15 minutes) and dedicated (over 15 minutes).
As a result, we have time series with four features (morning, afternoon, evening and overnight) over 17 timestamps (weeks) for each student, where each entry is one of the three engagement classes.

Additionally, we also discretise exam scores by binning them into three academic achievement classes: low, middle and high.
Linear interpolation was used to estimate the one-third and two-third quantiles.
The class sizes are roughly uniform, but not exactly even across all datasets since we avoid splitting ties at bin edges.

\subsection{Applying the CAPS framework}

% As simulating learner behaviour by LLMs is an emerging area of research \cite{Xu2025simulacra,Tao2025imperfection,Mannekote2025authoring}

\subsubsection{Step 0: pre-train M1}

Since public data applicable to our setting is not available, we utilise large language models (LLMs) to simulate realistic data that match the schema of the private data and use the generated data as \emph{surrogate} public data \cite{Hod2025surrogate}. 
Building on the approach of Hod et al.\ \cite{Hod2025surrogate}, we developed a prompt to generate a Python script to simulate learning habits time series with the same schema of our datasets.
Then the prompt is fed to \verb|Gemini-2.5-pro|, \verb|GPT-o3| and \verb|GPT-o4-mini-high| to generate ten scripts for the first and five scripts for each of the latter models\footnote{The prompt, generated scripts and source code for the subsequent experiments are available at \url{https://github.com/hibiki-i/CAPS}}.
Each script is run to generate 10,000 examples, summing up to 200,000 examples in total.
Finally, 100,000 data points are sampled from this pool uniformly at random to form our surrogate public dataset $\gX_\pub$.

Following Kingma et al.\ \cite{Kingma2014SSL}, the posterior for M1---of which the exact distribution is intractable---is approximated as follows:
\begin{equation}
    \quad p_{\thetab_1} (\z_1 \mid \x) \approx q_{\phib_1} (\z_1 \mid \x) = \gN(\z_1; \mub_{\phib_1}(\x), \diag(\sigmab_{\phib_1}(\x)^2) ),
\end{equation}
where $\z_1$ is a 16 dimensional latent variable.
We instantiate an M1 model by using 1D convolutional layers for the encoder $\phib_1$ and decoder $\thetab_1$ with ReLU activation based on the prior work by Desai et al.\ \cite{Desai2021timeVAE}.
To train M1, we employ $\beta$-VAE \cite{Higgins2017betaVAE}:
\begin{equation}
     \min_{\thetab_1, \phib_1} - \sE_{q_{\phib_1}(\z_1 \mid \x)} \left[\log p_{\thetab_1}(\x \mid \z_1)\right] + \beta_1 \KL\left(q_{\phib_1}(\z_1 \mid \x) \parallel p_{\thetab_1}(\z_1)\right).
\end{equation}
This helps avoid vanishing the KL term, a common issue known as \emph{posterior collapse} \cite{van-den-Oord2017VQVAE}, and disentangle latent representations \cite{Burgess2018disentangle}.
We set $\beta_1=10^{-3}$ throughout our experiments.
Moreover, for both M1 and M2, cyclical $\beta$-annealing \cite{Fu2019annealing} is implemented to improve training.

\subsubsection{Step 1: train M2 for cycle $t$}

We first prepare an unlabelled dataset $D_t^\prime$ consisting of 10,000 points generated from the pre-trained M1.
Then for M2 we use a standard VAE with the encoder and decoder being fully connected neural networks with ReLU activation and add a linear classifier for classifying $\z_1$ as in Kingam et al. \cite{Kingma2014SSL}:
\begin{align}
    \quad p_{\thetab_2} (\z_2 \mid y, \z_1) &\approx q_{\phib_2} (\z_2 \mid y, \z_1) \\
    &= \gN(\z_2; \mub_{\phib_2}(y, \z_1), \diag(\sigmab_{\phib_2}(y, \z_1)^2) ),\\
    \quad p_{\thetab_2}(y \mid \z_1) &\approx q_{\phib_2} (y \mid \z_1) = \Cat(y \mid \pib_{\phib_2}(\z_1) ),
\end{align}
where $\z_2$ is a 4 dimensional latent variable.
See \cref{sec:implementation} for the details of the model architecture.

To train M2, we have different loss functions for labelled and unlabelled data points:
\begin{align}
    &\begin{aligned}
        \text{Labelled:} \quad \gL(\z_1, y) = - \sE_{q_{\phib_2}(\z_2 \mid y, \z_1)}\left[\log p_{\thetab_2}(\z_1\mid y, \z_2) \right] + \\
        \beta_2 \KL\left(q_{\phib_2}(\z_2 \mid y, \z_1) \parallel p_{\thetab_2}(\z_2) \right)
    \end{aligned}\\
    &\begin{aligned}
    \text{Unlabelled:} \quad \gU_t(\z_1) = \sum_{y \in \sY_t} q_{\phib_2}(y \mid \z_1) \gL(\z_1, y) + \gH (q_{\phib_2}(y \mid \z_1)),
    \end{aligned}
\end{align}
where $\gH$ denotes the Shannon entropy and we assume that the prior of the label space $\sY_t$ is a uniform distribution in our case.
As recommended by Kingma et al.\ \cite{Kingma2014SSL}, we include a classification loss of $q_{\phib_2}(y \mid \z_1)$, so the objective of M1 becomes for some $\alpha$:
\begin{multline}
    \min_{\thetab_2, \phib_2} \sum_{(\x,y) \in D_t} \gL(\Mone(\x), y) + \sum_{\x \in D_t^\prime} \gU_t(\Mone(\x)) \\
    + \alpha \sE_{(\x,y) \in D_t}\left[ - \log q_{\phib_2}(y \mid \Mone(\x))\right],
\end{multline}
where $\Mone(\x)=\z_1$ denotes the latent features inferred by the frozen M1.
We set $\alpha=1$ and $\beta_2=10^{-3}$ throughout the experiments.
We also perform hyperparameter optimisation once for training M1 and M2 using Optuna \cite{Akiba2019optuna}, and the same hyperparameters are used in all stages (see \cref{sec:implementation} for more details).

We implement SPSSL based on the DP stochastic gradient descent (DP-SGD) mechanism \cite{Abadi2016DP-SGD} using the Opacus library \cite{Yousefpour2021opacus}.
Specifically, we use the Adam optimiser \cite{Kingma2015adam} instead of the standard SGD as recent research suggests that DP-Adam performs better than DP-SGD for VAE \cite{Ha2025DP-VAE}.
The SPSSL algorithm is described \cref{alg:semi-private}.

\begin{algorithm*}[ht]
\caption{Semi-private semi-supervised Adam for training M2}\label{alg:semi-private}
\begin{algorithmic}[1]
\REQUIRE Unlabelled public dataset $D_t^\prime$ and labelled private dataset $D_t$ of size $N_\priv$ for cycle $t$, private batch size $B_\priv$, public batch size $B^\pub$, step count $K$, clipping norm $C$, noise multiplier $\sigma$, learning rate $\gamma$, decay rates $\rho_1,\rho_2$, stability constant $\epsilon$
\STATE $\Theta_0 \gets 0$ \COMMENT{initialise parameters}
\STATE $\m_0 \gets 0$ \COMMENT{first moment}; $\vb_0 \gets 0$ \COMMENT{second moment}
\FOR{$k = 1, \dots, K$}
\STATE \text{Take a private mini-batch $B_k^\priv$ from $D_t$ with sample rate $B^\priv/N_\priv$}
\STATE \text{Calculate per-example gradients $\tilde{\g}_{k,j}^\priv$ for each $(\x_j^\priv, y_j^\priv)\in B_k^\priv$}
\STATE $\bar{\g}_{k,j}^\priv \gets \g_{k,j}^\priv / \max(1,\ \| \g_{k,j}^\priv \|_2 / C)$  \COMMENT{Clip gradients}
\STATE $\tilde{\g}_k^\priv \gets \frac{1}{B^\priv}\left( \sum_j \bar{\g}_{k,j}^\priv + \gN(0, \sigma^2 C^2 I) \right)$  \COMMENT{Add Gaussian noise}
\STATE \text{Take a public mini-batch $B_k^\pub$ of size $B^\pub$ from $D_t^\prime$ at random}
\STATE \text{Calculate the gradient $\g_k^\pub$ for $B_k^\pub$}
\STATE $\g_k \gets \tilde{\g}_k^\priv + \g_k^\pub$  
\STATE $\m_k \gets \rho_1 \m_{k-1} + (1 - \rho_1) \g_k;\; \vb_k \gets \rho_2 \vb_{k-1} + (1 - \rho_2) \g_k^2$
\STATE $\widehat{\m_k} \gets \m_k / (1 - \rho_1^k);\; \widehat{\vb_k} \gets \vb_k / (1 - \rho_2^k)$
\STATE $\Theta_k \gets \Theta_{k-1} - \gamma \widehat{\m_k} / (\sqrt{\widehat{\vb_k}} + \epsilon)$
\ENDFOR
\end{algorithmic}
\end{algorithm*}

\subsubsection{Step 2: update M1}

We employ the generative replay method \cite{Shin2017genReplay}, a simple yet powerful continual learning technique, for updating M1. 
Specifically, 10,000 unlabelled data points are generated from each of the M1+M2 stacked model and the M1 pre-trained (i.e.\ the replay ratio is 0.5).
Then the M1 is trained on these data randomly mixed by the non-DP Adam optimiser.

\section{Case study: results}\label{sec:results}

\subsection{Privacy accounting}

We used R\`enyi DP (RDP) \cite{Mironov2017RDP}, a stable and established method for privacy accounting, to calculate sufficient noise multipliers for target DP guarantee.
In addition, we also report accounting results by Gaussian DP (GDP) \cite{Dong2022GDP} based on recent recommendation by Gomez et al.\ \cite{Gomez2025numGDP}.
We do not account for privacy loss from hyperparameter optimisation, following a convention in prior DP research \cite{Tobaben2023few-shot,De2022DPscale}.

In \cref{tab:privacy-accounting}, $\mu$ is the parameter of GDP, and $\varepsilon$ is calculated by setting $\delta=10^{-3}$.
Regret $\Delta$ is a metric that quantifies the fit of GDP to the full privacy profile \cite{Kaissis2024regret}, and $\Delta < 10^{-2}$ is considered to well capture the privacy guarantee \cite{Gomez2025numGDP}, which is satisfied in all of our cases.
Since noise multipliers are calculated through RDP for target ($\varepsilon, \delta$), the accounting results show that the amount of noise may be too pessimistic for the privacy guarantee.

\begin{table}[tb]
    \centering
    \caption{Privacy accounting results. $\mu$ is the parameter of GDP, and $\Delta$ (regret) quantifies the fit of GDP to the full privacy profile.}
    \label{tab:privacy-accounting}
    \begin{tabular}{@{}rrrr@{}}
        \toprule
        $\varepsilon$ (RDP) & $\varepsilon$ (GDP) & $\mu$ & $\Delta$ (regret) \\
        \midrule
        1.0 & 0.83 & 0.35 & $0.43\cdot 10^{-2}$ \\
        2.0 & 1.75 & 0.63 & $0.24\cdot 10^{-2}$ \\
        4.0 & 3.49 & 1.12 & $0.96\cdot 10^{-2}$ \\
        \bottomrule
    \end{tabular}
\end{table}

\subsection{Utility of generative models}

To evaluate the utility of the generative models in downstream tasks, we employ academic achievement prediction performance as an indicator in this case study.
Note that, instead of training a classification model on synthetic data, we may use M2's classification functionality given by $q_{\phib_2}(y \mid \z_1)$.
To increase the number of samples, in addition to the real chronological order ($2022 \rightarrow 2023 \rightarrow 2024$), we included \emph{mock} orders (e.g.\ $2024 \rightarrow 2023 \rightarrow 2022$) and ran each experiment over 5 random seeds, summing up to $3! \cdot 5 = 30$ total runs.

\begin{figure}[tb]
    \centering
    \begin{subfigure}{0.5\linewidth}
        \centering
        \includegraphics[width=\linewidth]{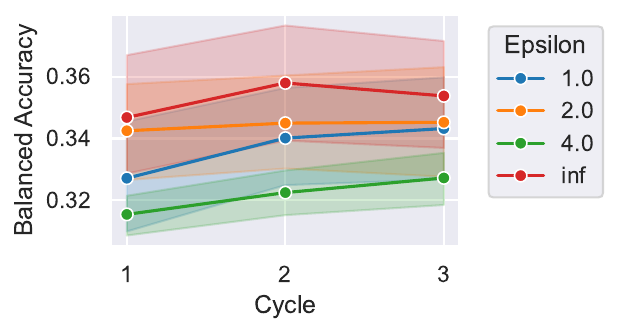}
        \Description{Line plots show the balanced accuracy (y-axis) over three cycles (x-axis) for epsilons 1.0, 2.0, 4.0 and infinity. They are slightly increasing over cycles.}
        \caption{Balanced accuracy}
        \label{fig:accuracy}
    \end{subfigure}\hfill
    \begin{subfigure}{0.5\linewidth}
        \centering
        \includegraphics[width=\linewidth]{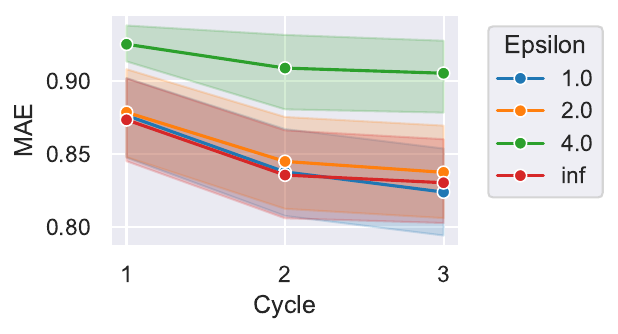}
        \Description{Line plots show the mean absolute error (y-axis) over three cycles (x-axis) for epsilons 1.0, 2.0, 4.0 and infinity. They are slightly decreasing over cycles.}
        \caption{Mean absolute error (MAE)}
        \label{fig:mae}
    \end{subfigure}
    \caption{Performance of generative models in academic achievement prediction for different privacy parameters and cycles within the CAPS framework. The shaded areas indicate 95\% confidence intervals. $\varepsilon=\infty$ is the non-DP baseline.}
    \label{fig:utility}
\end{figure}

\cref{fig:utility} shows test balanced accuracy and mean absolute error.
For example, test data for the classifier trained on the data of year 2022 consist of the data of year 2023 and 2024.
We observe that performance is mostly increasing over cycles for both metrics.
This indicates that CAPS effectively adapt the model over cycles, outperforming the one-shot baseline (i.e.\ the initial cycle).
Nonetheless, it should be noted that, as the baseline accuracy of the random guess classifier is $1/3$, none of the models exhibit practically feasible performance.
Indeed, unclear privacy-utility trade-off and larger uncertainty confirm the inherent difficulty in predicting academic achievement from learning habits.

\subsection{Quality of synthetic data}

To complement the above utility assessment, we evaluate the quality of generated data.
While there are a growing number of metrics for evaluating the quality of synthetic data such as fidelity and diversity \cite{Stenger2024evaluation}, R\"ais\"a et al.\ \cite{Raisa2025metrics} recently demonstrated that those metrics currently available are not consistent across different occasions and potentially provide misleading pictures. 
Thus, in the following we only rely on general statistical divergence metrics, and the results should be seen as one indicator among others that give nuanced understanding of generated data quality, leaving more rigorous real-world assessment for future work.

In particular, we define the following average Jensen-Shannon (AJS) divergence similarly to prior works \cite{Stenger2024evaluation,Li2022AJS,Ouyang2018AJS}.
For each time series $\x$ of 4 features (morning, afternoon, evening and overnight), let $f(\x)$ be a $4\cdot 4=16$ dimensional vector containing the median, mean, standard deviation and entropy of each feature.
Then the AJS divergence between a real dataset $D_t$ and a synthetic dataset $D_t^\syn$ for cycle $t$ is given as
\begin{align}
    \AJS(D_t, D_t^\syn) =& \frac{1}{3} \sum_{c=1}^3 \left( \frac{1}{16} \sum_{h=1}^{16} \JS (\widehat{P}_{t,c}^{(h)}, \widehat{Q}_{t,c}^{(h)} ) \right) \\\label{eq:ajs}
    \widehat{P}_{t,c}^{(h)} =& \left\{f_h(\mathbf{x}) \mid (\mathbf{x},y)\in D_t y=c\right\}, \\
    \widehat{Q}_{t,c}^{(h)} =& \left\{f_h(\mathbf{x}) \mid (\mathbf{x},y)\in D_t^{\mathrm{syn}}, y=c\right\}.
\end{align}
where $\JS$ denotes Jensen-Shannon divergence between two empirical distributions, $f_h(\x)$ is the $h$-th dimension of the vector $f(\x)$ and $c=1,2,3$ are the academic achievement classes.

\begin{figure}[tb]
    \centering
    \begin{subfigure}{0.5\linewidth}
        \centering
        \includegraphics[trim={0 0 0 0.7cm},clip,width=\linewidth]{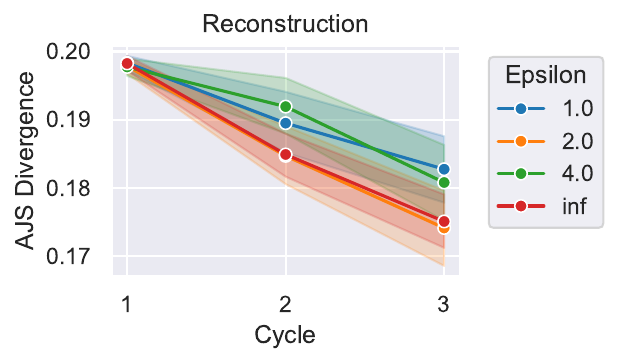}
        \Description{Line plots show the AJS divergence (y-axis) over three cycles (x-axis) for epsilons 1.0, 2.0, 4.0 and infinity. They are all slightly decreasing.}
        \caption{Reconstruction}
        \label{fig:ajs_recon}
    \end{subfigure}\hfill
    \begin{subfigure}{0.5\linewidth}
        \centering
        \includegraphics[trim={0 0 0 0.7cm},clip,width=\linewidth]{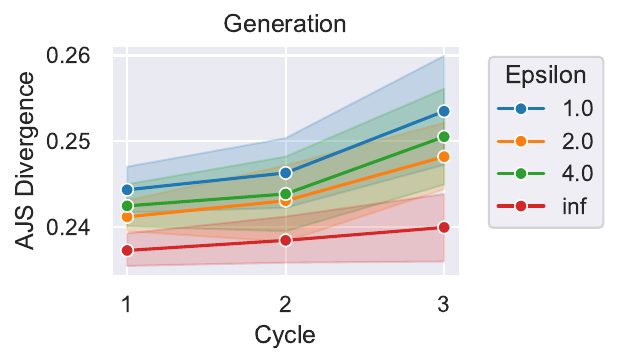}
        \Description{Line plots show the AJS divergence (y-axis) over three cycles (x-axis) for epsilon 1.0, 2.0, 4.0 and infinity. They are slightly increasing over cycles for, especially for smaller epsilons.}
        \caption{Generation from the prior}
        \label{fig:ajs_gen}
    \end{subfigure}
    \caption{AJS divergence defined in \cref{eq:ajs} between real and synthetic data (a) reconstructed from the real data and (b) conditionally generated by sampling from the prior. The shaded areas indicate 95\% confidence intervals. $\varepsilon=\infty$ is the non-DP baseline.}
    \label{fig:ajs}
\end{figure}

\cref{fig:ajs} shows the AJS divergence between real and reconstructed data (\cref{fig:ajs_recon}) as well as between real and synthetic data conditionally generated from prior samples $(\z_2, y)$ (\cref{fig:ajs_gen}).
We observe that the AJS divergence for reconstruction clearly decreases over cycles in our CAPS framework, while conditional generation is slightly degrading over cycles as the divergence is growing.
The former result is expected and confirms the effectiveness the CAPS framework in terms of learning the statical properties of real data over cycles, while the latter contradicts our hypothesis that the quality of private synthesis iteratively improves.
This seems to suggest that some \emph{bias} in the one-shot setting of the first cycle is amplified in the subsequent cycles.
The bias might come from LLM-generated training data or/and the training algorithm.
Moreover, this bias is larger for stronger DP protection (smaller $\epsilon$).
A potential explanation is that the mismatch between the prior $p_{\thetab_2}(\z_2)$ and the variational posterior $q_{\phib_2}(\z_2)$ of M1 at cycle 1 causes this issue \cite{Hoffman2016ELBO}.
This mismatch would introduce some bias in $\gX_1^\prime$ which is used for updating M1.
Then the updated M1 generates biased $D_1^\prime$ used to train M2 at cycle 2.
Since smaller $\epsilon$ adds more noise to learning from private data, at cycle 2, M2 learns more signal from the biased $D_1^\prime$, potentially proliferating the posterior-prior mismatch.
While this is a tentative, hypothetical explanation, we term this phenomenon as \emph{compounding bias effect} and leave more thorough investigation for future work.

\section{Discussion and conclusion}

\subsection{Discussion}

Despite the growing amount of RWD in education, concerns over data privacy limit access and have hindered data sharing in LA research, undermining the practice of open science and the progress of LA.
Although private synthesis is a promising approach to sharing sensitive data, its potential for RWD---often small in sample size and high in dimensionality---has been under-explored.
Notably, since sharing educational RWD is a continual process rather than a one-off event, merely applying existing methods falls short. 
Thus it is imperative to consider the domain’s specific characteristics and employ them to adapt existing private synthesis techniques.

The proposed CAPS framework advances this goal by drawing specifically on the iterative nature of educational practice.
It not only customises private synthesis methods for educational contexts and extends the availability of RWD, but also enables DBR by cyclically providing LA researchers with RWD.
While traditional control-group experiments such as randomised controlled trials offer reliable evidence, they are costly and often difficult to conduct because of ethical concerns in education \cite{Okumura2026RWE}.
Thus, DBR is essential for systematically improving educational practice while simultaneously supporting the discovery of RWE and theory development \cite{Barab2014DBR}.
In particular, LA plays a pivotal role by providing practical solutions within DBR \cite{Reimann2016LA-DBR}.
CAPS opens this landscape by iteratively sharing RWD in a privacy-preserving manner, thereby significantly increasing the impact of LA.

We further evaluated CAPS using authentic RWD in education.
Such evaluation is critical because open datasets, while readily available, rarely reflect the actual distributional characteristics of sensitive RWD.
Similar concerns have been raised within the DP community, where there is growing recognition of the need to evaluate DP machine learning techniques on sensitive datasets rather than solely on public benchmarks \cite{Tramer2024public}.
As a result, our case study extends its contribution beyond LA, offering insights that advance the broader DP research agenda.

The experimental results bring us several implications.
First, we relied on plain RDP to determine required noise to satisfy pre-defined DP guarantees owing to the stability of the underlying software.
As confirmed by our experiments, the standard RDP tends to overestimate privacy parameters for DP-SGD \cite{Dong2022GDP}.
Since privacy accounting is a rapidly evolving research area, a careful choice is needed in deployment.
In addition, it was demonstrated that the model utility improves over cycles in terms of downstream classification performance.
This suggests that the model adapts to learn latent features for different classes over time, effectively leveraging the synthetic data from earlier cycles.
The improvement in the model's reconstruction capability further supports effective cyclic adaption.
These findings suggest that CAPS effectively enables private synthesis in the context of iterative sharing of educational RWD, outperforming the traditional one-shot baseline.
Nonetheless, academic achievement prediction from learning habits may not be a practically feasible downstream task, and the compounding bias effect observed in the quality assessment suggests the need for further investigation on potential challenges.

Another important direction of subsequent research is real-world utility assessment of DP synthetic data.
In this case study, we only considered limited utility and quality evaluation, relying on statistical measures.
However, what practitioners care about most when using private synthesis is \emph{epistemic parity} \cite{Rosenblatt2023parity}.
Namely, an essential practical concern is whether the findings from downstream analyses on DP synthetic data are replicable on real data.
This must be assessed through real-world use cases of DP synthetic data, rather than statistical metrics alone.
Nonetheless, real-world assessment of epistemic parity is lacking not only in LA but also in DP literature \cite{Rosenblatt2023parity}.
Consequently, research on privacy-preserving sharing of RWD should be advanced by developing real-world assessment methods alongside methodological exploration of private synthesis.
This is particularly crucial for the development of LA infrastructures since DP-SGD introduces additional computational costs by calculating per-example gradients \cite{Ponomareva2023DPfy,Abadi2016DP-SGD}.
Investing in such expensive LA infrastructures that enable private synthesis will be challenging without evidence of real-world utility.

Finally, we must heed the caution over the use of large pre-trained models for DP tasks raised by Tram\`er et al.\ \cite{Tramer2024public}: large web-scraped data used for pre-training foundation models like LLMs contain personally identifiable information that was not intentionally shared for that purpose.
The recent work by Hong et al.\ \cite{Hong2025public} also raises concerns about legal implications of using web-scraped data for foundation models.
As discussed by Hod et al.\ \cite{Hod2025surrogate}, the use of LLM-simulated data as surrogate public data assumes that the training data of the LLMs are public with respect to training a model on the private data in question.
That is, our CAPS framework provides a DP guarantee only for RWD $D_t$, and LLM-simulated data are \emph{public} with respect to private synthesis of $D_t$, necessarily assuming that they are non-sensitive.
Since CAPS relies on public pre-training to handle the small-sample and high-dimensional RWD yet suitable public data are rarely available in education, careful ethical considerations are essential when using LLMs for CAPS.
As the survey by Viberg et al.\ \cite{Viberg2022definition} shows that definitions of privacy widely vary---or are sometimes absent altogether---in the LA literature, further discussion of the meaning of privacy-preserving data sharing and its ethical implications is needed within LA.

\subsection{Limitation}

An inherent assumption in CAPS is that the feature space remains identical, or at least similar, so that the pre-trained M1 can be shared with no or minimal architectural modification across cycles.
This requires consistent data collection and feature engineering throughout those cycles.
Our case study is also limited to a simplified setting of learning-habits RWD.
While the small sample size used in the experiments is intentional, this introduces a lack of diversity in underlying distributions.
The effectiveness of CAPS on other types of RWD and more diverse populations should be rigorously tested in future work.

Additionally, the quality assessment of conditionally generated synthetic data from prior samples---which is typically shared---reveals a potential challenge of the compounding bias effect.
While our metric is just one general indicator, this effect might influence downstream tasks on shared data in practice.
We offered a tentative explanation of the phenomenon, but further research on understanding and mitigating it is needed.
Particularly, since we often need to rely on LLM-generated data due to lack of public data, bias introduced by LLMs would require further investigation.
For example, if the prior-posterior mismatch is the root cause, cyclic adaption of not only M1 but also the prior $p(\z_2)$ would be worth exploring \cite{Hoffman2016ELBO}.

\subsection{Conclusion}

To address the lack of research on private synthesis of RWD in education, we proposed the CAPS framework and tested it on authentic RWD. 
Drawing on the iterative nature of educational practice, CAPS leverages public pre-training and cyclic adaption of a feature extractor, enabling iterative sharing of RWD in education.
As a result, it advances the practice of open science in LA and provides rich opportunities for DBR, thereby significantly increasing the impact of LA. 
The case study demonstrated the framework’s effectiveness, though closer examination also revealed potential challenges that warrant further investigation. 
Overall, this paper takes an essential first step towards sharing RWD in education and thereby significantly increasing the impact of LA.

\begin{acks}
This work was supported by CSTI SIP Grant Number JPJ012347 and JSPS KAKENHI Grant Numbers 23H00505, 25KJ1515.
\end{acks}

\bibliographystyle{ACM-Reference-Format}
\bibliography{references}

\appendix

% \section{LLM prompt}\label{sec:prompts}

% The following prompt was fed to LLMs to generate Python scripts that simulate the learning habits data used in the experiment.

% \begin{prompt}\ttfamily\small
%     You are a professional synthetic-data generator. \\
    
%     TASK \\
%     Write a python script to generate a realistic dataset of lower-secondary-school students’ self-practice behaviour during 17 weeks of mathematics study. \\
    
%     DATA-SHAPE \\
%     • N students  \\
%     • 17 weeks (week = 1…17)  \\
%     • 4 daily attributes: morning (5:00–11:59), afternoon (12:00–16:59), evening (17:00–23:59), overnight (00:00–04:59) \\
%     • Categorical values: 0 = inactive, 1 = light engagement, 2 = heavy engagement \\
%     • Data shape: (N, 17, 4) \\
    
%     STATISTICS \\
%     • Overall zero-inflation \\
%     • Include a realistic number of edge-case students \\
%     • Assume no holidays, breaks or exam periods \\
%     • Ensure a realistic    distribution that reflects 12 to 13 year-old students' behaviour, assuming that their task is voluntary and not graded \\
    
%     OUTPUT \\
%     • \detokenize{`generate.py` — Python script to generate the dataset. The script should generate a numpy array with shape (N, 17, 4) and save it to <path_to_save_file> with <random_seed> by running `python generate.py --save_path <path_to_save_file> --num_samples N --seed <random_seed>`}. \\
    
%     DELIVERY RULES \\
%     • Do not print any explanatory text, previews, or stats
% \end{prompt}

\section{Model architecture and hyperparameters}\label{sec:implementation}

For M1, both the encoder and decoder are 1D convolutional networks with two hidden layers of sizes 32 and 64. 
No hidden layers are set for M2. 
To mitigate overfitting, for both M1 and M2, dropout was implemented with probabilities 0.2 and 0.5 for the encoders and decoders, respectively. 
This also helps to avoid posterior collapse as strong decoders tend to ignore priors.
Adam optimiser was used for training both M1 and M2 with decay rates $\rho_1=0.9$ and $\rho_2=0.999$ and constant $\epsilon=10^{-8}$.
M1 was trained over 50 epochs.
Other hyperparameters are optimised by tree-structured Parzen estimator (TPE) algorithm \cite{Bergstra2011TPE} for 20 trials within the ranges shown in \cref{tab:hypers}.
For M2, hyperparameters optimisation was performed using data for year 2022 as a training set and data for 2023 as a holdout set.

\begin{table}[hb]
    \centering
    \caption{Ranges for hyperparameter optimisation}
    \label{tab:hypers}
    \begin{tabular}{@{}cll@{}}
        \toprule
        \multirow{2}{1.5em}{M1} & learning rate $\gamma$ & $[10^{-5}, 10^{-2}]$ (log-scale) \\
        & batch size & $[16, 512]$ ($\text{step}=8$) \\
        \midrule
        \multirow{5}{1.5em}{M2} & learning rate $\gamma$ & $[10^{-5}, 10^{-2}]$ (log-scale) \\
        & public batch size $B_\pub$ & $[16, 512]$ ($\text{step}=8$) \\
        & private batch size $B_\priv$ & $[1, |D_t|-1]$ ($\text{step}=1$) \\
        & epochs & $[1, 100]$ ($\text{step}=1$) \\
        & clipping norm $C$ & $[0.1, 5.0]$ ($\text{step}=0.1$) \\
        \bottomrule
    \end{tabular}
\end{table}

\end{document}